\documentclass[a4paper]{jpconf}
\usepackage{graphicx}
\usepackage{epstopdf}
\usepackage{amssymb}
\usepackage{amsmath}
\usepackage{hyperref}
\usepackage{xcolor}
\usepackage{floatflt}

\newcommand{\mm}[1]{\texttt{\textbf{#1}}}
\newcommand{\mme}[1]{\textnormal{\textit{#1}}}
\newcommand{\LR}{\texttt{LiteRed}~}
\begin{document}
\title{\texttt{LiteRed} 1.4: a powerful tool for the reduction of the multiloop integrals}

\author{Roman N. Lee}
\address{Budker Institute of Nuclear Physics, 630090, Novosibirsk, Russia}
\ead{R.N.Lee@inp.nsk.su}

\begin{abstract}
We review the \emph{Mathematica} package \texttt{LiteRed}, version $1.4$.
\end{abstract}

\section{Introduction}

Multiloop integrals are the building blocks of many calculations of radiative corrections in QFT. One of the key approaches to the calculation of the multiloop integrals is the IBP reduction, based on the integration-by-part identities between integrals \cite{ChetTka1981,Tkachov1981}. The IBP reduction almost necessarily should be done with the aid of various computer programs. One of the most successful methods of the IBP reduction is the Laporta algorithm \cite{Laporta2000}. The algorithm is easy to implement and to use, and allows for a number of programming improvements. These advantages explain why many modern most powerful reduction programs heavily rely on this algorithm, in particular,\texttt{ AIR} \cite{AnastasiouLazopoulos2004}, \texttt{FIRE} \cite{Smirnov2008}, \texttt{Reduze} \cite{Studeru2010,ManteuffelStuderus2012}, and many private versions. 

However, the Laporta algorithm has some weak points, which may put restrictions on its application. In particular, being intrinsically a brute-force search, this algorithm is both time- and memory-consuming. Another approach to the reduction is a derivation of the symbolic reduction rules. Its advantages are obvious: nothing is being solved in the process of reduction, therefore, the reduction is very fast. Symbolic rules are small in size, so, they can be easily saved for future calculations. The bottleneck of this approach is the search of these symbolic rules. Much effort has been devoted to the developement of the approach connected with the notion of the Groebner basis \cite{Tarasov1998,Gerdt2005a,SmirSmi2006a,Smirnov:2006tz}, but, for now, this approach is far from being satisfactory. Probably, the only, partly successful, attempt to implement this approach has been made in \texttt{FIRE}, where the notion of s-bases \cite{SmirSmi2006a,Smirnov:2006tz} has been used.

Recently, the new IBP reduction package \LR has been presented in Ref. \cite{Lee2012}. This package uses completely different approach to the reduction. At first stage it tries to find symbolic reduction rules using heuristics. Then it applies the rules to the specific reduction. Found rules are very lightweight and can be easily stored for the reusage. In this contribution we describe the version $1.4$ of the \LR package. The package can be downloaded from \href{http://www.inp.nsk.su/~lee/programs/LiteRed/}{\texttt{http://www.inp.nsk.su/\textasciitilde{}lee/programs/LiteRed/}}.


\section{Multiloop integral}

To fix notation, let us assume that we are interested in the calculation of the $L$-loop integral depending on the $E$ external momenta $p_{1},\ldots,p_{E}$. There are $N=L(L+1)/2+LE$ scalar products depending on the loop momenta $l_{i}$:
\begin{equation}
s_{ij}=l_{i}\cdot q_{j}\,,\ 1\leqslant i\leqslant L,j\leqslant L+E,
\end{equation}
where $q_{1,\ldots,L}=l_{1,\ldots,L}$, $q_{L+1,\ldots,L+E}=p_{1,\ldots,E}$.

The general form of the integral is the following
\begin{align}
J\left(\boldsymbol{n}\right)&=J(n_{1},n_{2},\ldots,n_{N})=\int\frac{d^{d}l_{1}\ldots d^{d}l_{L}}{D_{1}^{n_{1}}D_{2}^{n_{2}}\ldots D_{N}^{n_{N}}}\,,\label{eq:J}
\\
D_{\alpha}&=a_{\alpha}^{ij}l_{i}\cdot l_{j}+2b_{\alpha}^{ik}l_{i}\cdot p_{k}+c_{\alpha}\,.\nonumber
\end{align}
Here $a_\alpha$, $b_\alpha$, and  $c_\alpha$ are  $L\times L$ matrices, $L\times E$ matrices, and numbers, respectively.
As usual, we assume that $D_1,\ldots,D_N$ form a complete basis in the sense that any $s_{ik}$ can be uniquely expressed in terms of $D_{\alpha}$. The multiindex $\boldsymbol{n}=(n_1,\ldots,n_N)$ can be thought of as a point in $\mathbb{Z}^{N}$. Some of $D_{\alpha}$ correspond to the denominators of the propagators, others may correspond to the irreducible numerators. E.g., the $K$-legged $L$-loop diagram with generic external momenta corresponds to $E=K-1$ and the maximal number of denominators is $M=E+3L-2$, so that the rest $N-M=(L-1)(L+2E-4)/2$ functions correspond to irreducible numerators.

\subsection{Differential equations}
The differential equations can be used for finding the master integrals. The simplest type of such equations is the differential equation with respect to the mass. Probably, the first example of their application is presented in Refs. \cite{Kotikov1991,Kotikov1991a,Kotikov1991b}. The differential equations with respect to the invariant constructed of the external momenta have been introduced and applied in Refs. \cite{GehrRem2000,GehrRem2001,GehrRem2001a}. In general case, when there are $E>2$ external vectors, we have the following formulas:
\begin{align}
\frac{\partial}{\partial\left(p_{1}\cdot p_{2}\right)}J\left(\mathbf{n}\right) & =\sum\left[\mathbb{G}^{-1}\right]_{i2}p_{i}\cdot\partial_{p_{1}}J\left(\mathbf{n}\right)=\sum\left[\mathbb{G}^{-1}\right]_{i1}p_{i}\cdot\partial_{p_{2}}J\left(\mathbf{n}\right)\,,\nonumber \\
\frac{\partial}{\partial\left(p_{1}^{2}\right)}J\left(\mathbf{n}\right) & =\frac{1}{2}\sum\left[\mathbb{G}^{-1}\right]_{i1}p_{i}\cdot\partial_{p_{1}}J\left(\mathbf{n}\right)\,.\label{eq:Dinv}
\end{align}
where $\mathbb{G}=\mathbb{G}\left(p_{1},\dots,p_{E}\right)=\begin{pmatrix}p_{1}^{2} & \cdots & p_{1}\cdot p_{E}\\
\vdots & \ddots & \vdots\\
p_{1}\cdot p_{E} & \cdots & p_{E}^{2}
\end{pmatrix}$ is a Gram matrix.

Acting by the operator on the right-hand side on the integrand and
performing the IBP reduction, one obtains the differential equation
for $J\left(\mathbf{n}\right)$.

\subsection{Dimensional recurrences}

Probably, the first appearance of the dimension shifting relations is in Ref.  \cite{DerkachovHonkonenPismak1990}, where they have been derived for certain three-loop integrals in the momentum representation. Later, in Ref. \cite{Tarasov1996} Tarasov derived dimensional relations using the paramentric representation. It is interesting that the first approach led to the lowering recurrence, while the latter one led to the raising recurrence. The lowering (raising) recurrence relates one integral in $d+2$ ($d-2$) dimensions to several integrals in $d$ dimensions.

As it was shown in Ref. \cite{Lee2010a}, the two recurrences can be represented as
\begin{equation}
J^{\left(d-2\right)}\left(\mathbf{n}\right)=(\mu/2)^{L}\det\left\{ 2^{\delta_{ij}}\frac{\partial D_{k}}{\partial s_{ij}}A_{k}|_{i,j=1\ldots L}\right\} J^{\left(d\right)}\left(\mathbf{n}\right).\label{eq:rDRR}
\end{equation}
\begin{equation}
J^{\left(d+2\right)}\left(\mathbf{n}\right)=\frac{(2\mu)^{L}\left[V\left(p_{1},\ldots,p_{E}\right)\right]^{-1}}{\left(d-E-L+1\right)_{L}}P\left(B_{1},\ldots,B_{N}\right)J^{\left(d\right)}\left(\mathbf{n}\right),\label{eq:lDRR}
\end{equation}
where $\mu=\pm1$ for Euclidean/Minkovskian metric, $\alpha_{L}=\alpha\left(\alpha+1\right)\ldots\left(\alpha+L-1\right)$
is the Pochhammer symbol, $V\left(v_{1},\ldots v_{k}\right)=\det\mathbb{G}\left(v_{1},\ldots v_{k}\right)$
is the Gram determinant, and $P\left(D_{1},\ldots,D_{N}\right)=V(q_{1},\ldots q_{L+E})$.

The operators $A_{\alpha}$ and $B_{\alpha}$ are defined as follows 
\begin{align}
\left(A_{i}J^{\left(\mathcal{D}\right)}\right)\left(n_{1},\ldots,n_{N}\right) & =n_{i}J^{\left(\mathcal{D}\right)}\left(n_{1},\ldots,n_{i}+1,\ldots,n_{N}\right),\nonumber \\
\left(B_{i}J^{\left(\mathcal{D}\right)}\right)\left(n_{1},\ldots,n_{N}\right) & =J^{\left(\mathcal{D}\right)}\left(n_{1},\ldots,n_{i}-1,\ldots,n_{N}\right).
\end{align}

\section{Parametric representation}

The parametric representation (or Feynman parametrization), is, of no doubt, one of the most useful tools for the multiloop calculations. It is important for both analytical (in particular, with subsequent Mellin-Barnes representation)  and numerical (in particular, together with sector decomposition) calculations of the multiloop integrals. Moreover, it may serve as an fundamental definition of the multiloop integrals for the case of non-integer dimensionality $d$. 

But it is also important that the parametric representation is very useful for revealing relations between the integrals. One  example is Tarasov's derivation of the raising dimensional recurrence relation \cite{Tarasov1996}. In Ref. \cite{Pak2012} the algorithm, based on the use of parametric representation, for the identification of the master integrals has been introduced. In recent paper \cite{LeePomeransky2013} it was shown that parametric representation allows for a simple determination of the number of master integrals in the given sector. 

\LR uses parametric representation for two purposes. First, it finds equivalent simple sectors by comparing their parametric representation. To account for the possible permutation of the parameters, it uses an approach combining the ideas from Ref. \cite{Pak2012} and Ref. \cite{NicMeBa1977}. Second, it uses parametric representation to determine zero sectors. In this section we describe shortly this new algorithm.

The parametric representation of the integral $J\left(\mathbf{n}\right)$ has the form
\begin{equation}\label{eq:FP1}
J(\mathbf{n})=\frac{\Gamma\left(\Sigma n -Ld/2\right)}{\prod_{\alpha}\Gamma\left(n_{\alpha}\right)}\int\prod_{\alpha}dz_{\alpha}z_{\alpha}^{n_{\alpha}-1}\delta\left(1-\Sigma z\right)\frac{F^{Ld/2-\Sigma n}}{U^{\left(L+1\right)d/2-\Sigma n}}\,,
\end{equation}
where $\Sigma n=\sum_{\alpha}n_{\alpha}$, $\Sigma z=\sum_{\alpha}z_{\alpha}$, $U$ and $F$ are the homogeneous polynomials of degrees $L$ and $L+1$, respectively.
These polynomials can be expressed in terms of quantities 
\begin{equation}
a^{ij}=\sum_{\alpha}z_{\alpha}a_{\alpha}^{ij},\quad b^{i}=\sum_{\alpha}z_{\alpha}b_{\alpha}^{ij}p_{j},\quad c=\sum_{\alpha} z_{\alpha}c_{\alpha}
\end{equation}
as follows 
\begin{equation}
U=\det\left(a\right),\quad F=c\det\left(a\right) -\left(a^{\mathrm{Adj}}\right)^{ij}b^{i}\cdot b^{j},
\end{equation}
where $a^{\mathrm{Adj}}=\det\left(a\right)a^{-1}$ is the adjoint matrix. The representation \eqref{eq:FP1} does not make sense when some of $n_\alpha$ are nonpositive integers. In this case one has to replace the corresponding integration with the derivative at zero point. The resulting formula can be written as 
\begin{equation}\label{eq:FP}
J(\mathbf{n})=\Gamma\left(\Sigma n-Ld/2\right)
\prod_{\alpha}\hat{n}_\alpha
\delta\left(1-\Sigma_+ z\right)\frac{F^{Ld/2-\Sigma n}}{U^{\left(L+1\right)d/2-\Sigma n}}\,,
\end{equation}
where the functional $\hat{n}_\alpha$ is determined as 
\begin{equation}
\hat{n}_\alpha[\phi(z_\alpha)]=\left\{\begin{array}{rl}
\int_0^{\infty} \frac{dz_\alpha z_\alpha^{n_\alpha-1}}{\Gamma(n_\alpha)}\phi(z_\alpha) & n_\alpha>0\\
(-1)^{n_\alpha}\phi^{(-n_\alpha)}(0) & n_\alpha\leqslant 0
\end{array}\right.
\end{equation}
and the sum $\Sigma_+ z=\sum_{\alpha} \theta_\alpha z_\alpha$ ($\theta_\alpha=\Theta(n_\alpha-1/2)$) goes over the variables, corresponding to the denominators. Remarkably,  it is possible to rewrite Eq. \eqref{eq:FP} in the form,
which contains $U$ and $F$ only in the combination $F+U$. Similar to Ref.  \cite{LeePomeransky2013}, we have
\begin{equation}\label{eq:FPG}
J(\mathbf{n})=
\frac{\Gamma\left[d/2\right]}{\Gamma\left[\left(L+1\right)d/2-\Sigma n\right]}\prod_{\alpha}\hat{n}_\alpha G^{-d/2}\,,
\quad G=F+U\,.
\end{equation}

The scaleless integral can be defined as the one which gains additional non-unity factor under some linear transformation of the loop momenta. In dimensional regularization scaleless integrals are set to zero.  If $j\left(\theta_{1},\ldots,\theta_{N}\right)$ is scaleless, then all integrals of the sector $\left(\theta_{1},\ldots,\theta_{N}\right)$ are zero. We will call such a sector a \emph{zero sector}. 

A simple criterion of zero sectors has been formulated in Ref. \cite{Lee2008}. According to this criterion, the sector is zero if the solution of the IBP equations in the corner point $\left(\theta_{1},\ldots,\theta_{N}\right)$ result in the identity $ $$j\left(\theta_{1},\ldots,\theta_{N}\right)=0$. Note that this criterion may miss some scaleless sectors. Let us explain on a simple example why this happens. Consider the massless one-loop onshell propagator integral
\[
J\left(n_{1},n_{2}\right)=\int\frac{d^{d}l}{\left[l^{2}\right]^{n_{1}}\left[\left(l-k\right)^{2}\right]^{n_{2}}},\quad k^{2}=0\,.
\]

Obviously, this integral is zero for any $n_{1}$ and $n_{2}$. However,
it can be explicitely checked that the solution of the IBP identities in the corner point of the sector $\left(1,1\right)$ does not result directly to $J\left(1,1\right)=0$. In order to prove that the integral $J\left(1,1\right)$ is scaleless, let us consider instead the following operator
\[
O=\partial_{l}\cdot\left(l+\left(l\cdot k\right)\tilde{k}-\left(l\cdot\tilde{k}\right)k\right),
\]
where $\tilde{k}$ is an auxiliary vector chosen to satisfy the conditions $\tilde{k}^{2}=0$ and $\tilde{k}\cdot k=1$. It is easy to check that $Oj\left(1,1\right)=\left(d-4\right)j\left(1,1\right)$. Since the operator $O$ is a generator of the linear transformation $l\to l+\epsilon\left(l+\left(l\cdot k\right)\tilde{k}-\left(l\cdot\tilde{k}\right)k\right)$,
the integral $j\left(1,1\right)$ is scaleless. The reason why the
IBP identities failed to lead to the identity $J\left(1,1\right)=0$
is that the construction of this identity required introduction of
the auxiliary vector $\tilde{k}$. 

In some cases the number of zero sectors overlooked by the criterion of Ref. \cite{Lee2008} is rather big. So, we formulate below another criterion, based on parametric representation, which detects virtually all zero sectors.

As we said above, for the detection of zero sectors it is sufficient to consider only the integral in the corner point of the sector. In particular, we may set all $z_\beta$, corresponding to numerators, to zero. In what follows we assume this is done and $z_\alpha$ denotes a parameter, corresponding to the denominator of the sector.
Consider an infinitesimal scaling of these parameters
\begin{equation}\label{eq:subst1}
z_{\alpha}\to\tilde{z}_{\alpha}=\left(1+k_{\alpha}\omega\right)z_{\alpha}\,.
\end{equation}
Here $\omega$ is the infinitesimal parameter, and $k_{\alpha}$ are some finite coefficients. Suppose that we are able to find such $k_{\alpha}$ that the function $G$ scales as follows:
\begin{equation}\label{eq:Gscale}
G\left(\tilde{z}\right)=\left(1+\omega\right)G\left(z\right).
\end{equation}
Then, making the change \eqref{eq:subst1} in Eq. \eqref{eq:FPG},  we get
\begin{equation}\label{eq:Jscale}
J\left(\boldsymbol{\theta}\right)=\left[1+\omega\left(\sum_{\alpha}k_{\alpha}-d/2\right)\right]J\left(\boldsymbol{\theta}\right).
\end{equation}

The equation \eqref{eq:Gscale} does not depend on $d$, therefore, suitable $k_\alpha$, if they exist at all, can be chosen also independent of $d$. Therefore, the coefficient in Eq. \eqref{eq:Jscale} in front of $\omega$ is not zero, and the integral $J\left(\boldsymbol{\theta}\right)$ is scaleless. 

Note that Eq.  \eqref{eq:Gscale} can be cast as 
\begin{equation}\label{eq:k_equation}
\sum_{\alpha}k_{\alpha}z_{\alpha}\frac{\partial G(z)}{\partial z_{\alpha}}=G(z)\,,
\end{equation}
which should be understood as equality of the two polynomials of $z_\alpha$. Collecting the coefficients in front of distinct monomials, we obtain a linear system of equation with respect to $k_\alpha$. The existence of the solution of this system can be established by ordinary algebraic means.

Therefore, we get the following\\ 
\textbf{Criterion of zero sector:} \emph{For a given sector, construct $G=F+U$. The sector is zero if Eq.\eqref{eq:k_equation} has a $z$-independent solution with respect to $k_\alpha$.}

It is just this criterion which is implemented in \texttt{LiteRed1.4}.

%
%

\section{How \LR finds reduction rules}
\begin{figure}
\centering\includegraphics[width=5cm]{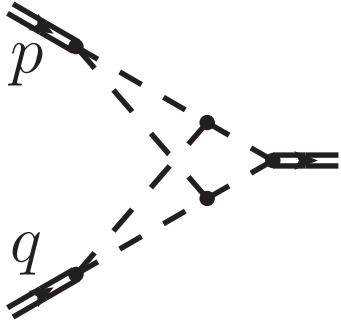}
\caption{Two-loop massless onshell vertex.}
\label{fig:1}
\end{figure}
When trying to find the complete set of the reduction rules,  \LR performs, roughly speaking, the same steps as a person would do. Let us explain this on the example of the two-loop massless onshell vertex shown in Fig. \ref{fig:1}.
We choose the basis
\begin{equation}
\left\{D_1,\ldots,D_7\right\}=\left\{l^2,r^2,(p-l)^2,(q-r)^2,(-l+p+r)^2,(l+q-r)^2,(l-r)^2\right\}\,.
\end{equation}
The function $D_7=(l-r)^2$ corresponds to the irreducible numerator. The diagram in Fig. \ref{fig:1} corresponds to the sector $(1,1,1,1,1,1,0)$ (see the definition of sectors in Ref. \cite{Lee2012}).
\begin{itemize}
\item First, \LR solves the IBP and LI identities in the general point $\mathbf{n}=(n_1,n_2,n_3,n_4,n_5,n_6,n_7)$ with respect to the most complex integrals. 
Then it shifts the indices in the rules found so that they all have the form $J(n_1,n_2,n_3,n_4,n_5,n_6,n_7)\to\ldots$
\item After that it analyzes the right-hand side to determine when each rule is applicable. The inapplicability may come from zeros in the denominators, or from the positive shifts in the indices, corresponding to the numerator, e.g., if the right-hand side contains the integral $J(n_1,n_2,n_3,n_4,n_5,n_6,n_7+1)$. Note that it often happens that the latter integral  appears with the coefficient $n_7$ and then does not result in the applicability condition $n_7\neq0$. Acting in this way, \LR finds 9 rules with the following applicability conditions: $\{c_1,\ldots,c_7\}=\{n_6\neq 1,n_5\neq 1,\neg \left(n_3=1\lor n_7=0\right),\neg \left(n_4=1\lor n_7=0\right),n_6\neq 1,\neg \left(n_1=1\lor n_7=0\right),\neg \left(n_6=1\lor n_7=0\right),\neg \left(n_2=1\lor n_7=0\right),\neg \left(n_5=1\lor n_7=0\right)\}$. 
\item None of the found rules is applicable when the condition $\neg (c_1\lor \ldots \lor c_7)$ is fulfilled. \LR reduces this condition to disjunctive normal form:
$\left(n_5=1\land n_6=1\land n_7=0\right)\lor \left(n_1=1\land n_2=1\land n_3=1\land n_4=1\land n_5=1\land n_6=1\right)$.
\item Then it takes the first alternative $\left(n_5=1\land n_6=1\land n_7=0\right)$ and tries to find the rules for the integral $J(n_1,n_2,n_3,n_4,1,1,0)$. It starts from the IBP and LI identities at the point $(n_1,n_2,n_3,n_4,1,1,0)$ and checks for the possibility to shift indices in order to reduce the found rules to the form $J(n_1,n_2,n_3,n_4,1,1,0)\to\ldots$. In contrast to all-indeterminate case, this is not always possible, so if the appropriate rule is not found, \LR starts to generate and solve identities in the points neighboring $(n_1,n_2,n_3,n_4,1,1,0)$. In fact, this search is the Laporta algorithm augmented by the procedure which checks for possible shifts.
\item When the appropriate rule is found, the condition of its applicability is constructed. In our case, this condition looks like $n_5=1\land n_6=1\land n_7=0\land n_4\neq 1$.
\item Performing the same steps, \LR finds 17 rules which reduce all integrals in the sector except $J(1,1,1,1,1,1,0)$ which it declares a master.
\end{itemize}

Note that \LR succesfully finds the reduction rules for much more complicated cases than the one presented above. In particular, it succeeds for the four-loop massless propagators and some other complicated cases.

\section{Short reference guide for \LR package}

The package is loaded by the command \mm{<<LiteRed`}. Let us describe briefly the most important procedures of \LR.

\begin{description}
\item[\mm{Declare[\mme{vars},\mme{type}]}] ---
variable declaration. Here \mme{vars} --- either variable name or the list of variable names, \mme{type} is either \mm{Vector} or \mm{Number}. Be sure to declare all variables that enter the basis definition (see below). Both vectors and numbers can be declared in one construct.\\
\emph{Example:} \mm{Declare[\{l,q\},Vector,mm,Number]}.

\item[\mm{NewBasis[\mme{name},\{\mme{D1,D2,\ldots}\},\mme{loopmoms},\mme{\color{gray}options}]}] --- definition of the basis. Here \mme{name} is the variable which will be the basis name. It will appear in many commands, associated with a given basis, like \mm{GenerateIBP[\mme{name}]}. Derived objects, like IBP identities, reduction rules, will be associated with this name. Make sure this name is not used anywhere else, but in the appropriate commands. \mm{\{\mme{D1,D2,\ldots}\}} is a list of functions $D_\alpha$. The scalar product is entered as \mm{sp[p1,p2]}. Third argument \mme{loopmoms} is a list of loop momenta.
The following options can be appended:
\begin{itemize}
\item \mm{Directory->"\mme{dirname}"} --- determine the directory, where all basis definitions will be saved.
\item \mm{Append->True} --- if the set of the functions $D_\alpha$, given in the first parameter, is not complete, append some automatically chosen numerators.
\item \mm{GenerateIBP->True} --- generate IBP identities, see the corresponding procedure below.
\item \mm{AnalyzeSectors->True} --- determine zero and simple sectors, see the corresponding procedure below.
\item \mm{FindSymmetries->True} --- find equivalent sectors, see the corresponding procedure below.
\end{itemize}
In case of success, \mm{NewBasis} creates objects 
\mm{Ds[\mme{name}]}, 
\mm{SPs[\mme{name}]},
\mm{LMs[\mme{name}]},
\mm{EMs[\mme{name}]}, and
\mm{Toj[\mme{name}]}. Their meaning is explained in the output of the \mm{NewBasis} procedure. The integrals of the basis are denoted as \mm{j[\mme{name},\mme{n1},\mme{n2},\ldots]}.

\emph{Example:} \mm{NewBasis[b1,\{sp[l,l]+mm,sp[l-q,l-q]\},\{l\},Directory->"bdir"]}\\
Convert the explicit expression to \mm{j} form with \mm{Toj}:\\
\mm{Toj[b1,sp[l,q](sp[l,l]+mm)\char`^{}{-1}(sp[l-q,l-q])\char`^{}-2]}\\
gives
\mm{j[b1,0,2]-j[b1,1,1]+(sp[q,q]-mm)j[b1,1,2]}. The inverse transformation can be done with \mm{Fromj[\mme{expr}]}.

\item[\mm{GenerateIBP[\mme{name}]}] generates IBP and LI identities for the basis. IBP identities in the point $(n_1,n2,\ldots)$ can be retrieved by \mm{IBP[\mme{name}][\mme{n1,n2,\ldots}]}.

\item[\mm{AnalyzeSectors[\mme{name},\mme{\color{gray}pattern}]}]  finds zero sectors and some other objects as it reports in the output. 
\mm{ZeroSectors[\mme{name}]} is a list of zero sectors, each element have the form \mm{js[\mme{name},$\theta_1$,$\theta_2$,\ldots]}.
The optional parameter \mme{pattern} tells the procedure to analyze only sectors matching pattern.  E.g. if the last two $D_\alpha$ correspond to irreducible numerators, use \mm{AnalyzeSectors[\mme{name},\{\char`_\char`_,0,0\}]}. This procedure should be called before the call of  \mm{FindSymmetries}.

\item[\mm{FindSymmetries[\mme{name}]}] finds equivalent sectors and forms. It generates several objects as it reports, including the list of unique sectors \mm{UniqueSectors[\mme{name}]} and the list \mm{MappedSectors[\mme{name}]}  of sectors which can be mapped onto unique ones. For each mapped sector \mm{js[\mme{name},$\theta_1$,$\theta_2$,\ldots]} the mapping rules can be retrieved as \mm{jRules[\mme{name},$\theta_1$,$\theta_2$,\ldots]}.

\item[\mm{SolvejSector[js[\mme{name},$\theta_1$,$\theta_2$,\ldots],\mme{\color{gray}options}]}] is a procedure which performs a heuristic search of the reduction rules for a given sector. If it succeeds, the list of found rules can be retrieved as \mm{jRules[\mme{name},$\theta_1$,$\theta_2$,\ldots]}. It returns the number of master integrals found. Typical usage is \mm{SolvejSector/@UniqueSectors[\mme{name}]}. The useful options include
\begin{itemize}
\item \mm{Depth -> \mme{n}} set heuristic search depth. Default is $n=2$
\item \mm{SR -> True} use internal symmetries of the sector.
\item \mm{TimeConstrained -> \mme{n}} set time constraint in seconds. 
\end{itemize}

\item[\mm{DiskSave[\mme{name}]}] save all definitions to disk (see option \mm{Directory} in \mm{NewBasis}).

\item[\mm{IBPReduce[\mme{expr}]}] performs the IBP reduction of the expression \mme{expr}.
\end{description}

Several additional tools are included in the package:
\begin{description}

\item \mm{Dinv[j[\mme{name},\mme{n1},\mme{n2},\ldots],sp[p,q]]} returns the derivative with respect to the invariant constructed of the external momenta.

\item \mm{RaisingDRR[\mme{name},\mme{n1},\mme{n2},\ldots]} returns the right-hand side of the dimensional recurrence relation \eqref{eq:rDRR}. Note that the factor $\mu^{L}=-1$ for Minkovskian
metrics and odd number of loops should be taken into account manually.

\item \mm{LoweringDRR[\mme{name},\mme{n1},\mme{n2},\ldots]} returns the right-hand side of the dimensional recurrence relation \eqref{eq:lDRR}.

\item \mm{FeynParUF[js[\mme{name},$\theta_1$,$\theta_2$,\ldots]]} returns the list \mm{\{$U$,$F$,\{x1,x2,\ldots\}}, where $U$ and $F$ are the functions  entering the parametric representation of the integrals in the given sector, and \mm{x1,x2,\ldots} are the parameters.
\end{description}

\paragraph{Learning more}
One is encouraged to examine the examples that are included in the distribution. Another good starting point to know
more about the functions of the package is to submit a command \texttt{\textbf{?LiteRed`{*}}}.

\section{Conclusion}

In this contribution we have reviewed the
\LR package performing the IBP reduction of the multiloop integrals. We have described a new 
algorithm of detecting the zero sectors implemented in \LR version 1.4.

\emph{Acknowledgments} 
I am grateful to the organizers of ACAT-2013 for the support and hospitality.
This work is supported by the Russian Foundation for Basic Research through grant 11-02-01196 and
by the Ministry of Education and Science of the Russian Federation. 

\section*{References}

\bibliographystyle{JHEP}
\bibliography{LiteRed14}

\end{document}